\documentclass[conference]{IEEEtran}
\IEEEoverridecommandlockouts
\newcommand{\mysplit}[1]{%
  \begin{tabular}[t]{@{}c@{}}
    #1
  \end{tabular}
  }

\IEEEoverridecommandlockouts 
\usepackage{graphicx}
\usepackage{balance}
\usepackage{algorithmic}
\usepackage{algorithm}
\usepackage{listings}
\usepackage[OT4,T1]{fontenc}
\usepackage[cmex10]{amsmath}
\usepackage{url}
\usepackage{multirow}

\usepackage{tikz}
\usepackage{textcomp}
\usepackage{hyperref}
\usepackage{lipsum}

\newcommand\copyrighttext{%
  \footnotesize \textcopyright 2012 IEEE. Personal use of this material is permitted.
  Permission from IEEE must be obtained for all other uses, in any current or future
  media, including reprinting/republishing this material for advertising or promotional
  purposes, creating new collective works, for resale or redistribution to servers or
  lists, or reuse of any copyrighted component of this work in other works.}
  
\newcommand\copyrightnotice{%
\begin{tikzpicture}[remember picture,overlay]
\node[anchor=south,yshift=10pt] at (current page.south) {\fbox{\parbox{\dimexpr\textwidth-\fboxsep-\fboxrule\relax}{\copyrighttext}}};
\end{tikzpicture}}

\title{Quantitative Impact of Label Noise on the Quality of Segmentation of Brain Tumors on MRI scans}
\author{
\IEEEauthorblockN{Micha{\l} Marcinkiewicz}
\IEEEauthorblockA{
Netguru \\
ul.\ Wojskowa 6, 60-792 Pozna\'n, Poland\\
Email: michal.marcinkiewicz@netguru.com}
\and
\IEEEauthorblockN{Grzegorz Mrukwa}
\IEEEauthorblockA{
Netguru \\
ul.\ Wojskowa 6, 60-792 Pozna\'n, Poland\\
Email: grzegorz.mrukwa@netguru.com}}

\newcommand{\Lim}[1]{\raisebox{0.5ex}{\scalebox{0.8}{$\displaystyle \lim_{#1}\;$}}}

\begin{document}
\suppressfloats
\maketitle              
\copyrightnotice

\begin{abstract}
Over the last few years, deep learning has proven to be a great solution to many problems, such as image or text classification. Recently, deep learning-based solutions have outperformed humans on selected benchmark datasets, yielding a promising future for scientific and real-world applications. Training of deep learning models requires vast amounts of high quality data to achieve such supreme performance. In real-world scenarios, obtaining a large, coherent, and properly labeled dataset is a challenging task. This is especially true in medical applications, where high-quality data and annotations are scarce and the number of expert annotators is limited. In this paper, we investigate the impact of corrupted ground-truth masks on the performance of a neural network for a brain tumor segmentation task. Our findings suggest that a) the performance degrades about 8\% less than it could be expected from simulations, b) a neural network learns the simulated biases of annotators, c) biases can be partially mitigated by using an inversely-biased dice loss function. 
\end{abstract}

\section{Introduction}
\IEEEoverridecommandlockouts\IEEEPARstart {T}{he} HUMAN brain is proficient in recognizing patterns in a variety of domains: visual, auditory, etc. Its performance is always treated as the golden standard for the assessment and a level to beat using machine learning (ML) and deep learning (DL) models. As it stands, datasets are labeled by human annotators, with different levels of training, predispositions, and of course, also harbor their own biases, which have an impact on the quality of their annotations. Reducing the errors in datasets, also called label noise, calls for double- and triple-checking (usually done by different annotators), which requires a vast amount of work.

For example, in the classification of natural images -- such as the ones included in the famous ImageNet dataset~\cite{imagenet_cvpr09} -- the human classification error rate was estimated at 5.1\% by Russakovsky et al.~\cite{Russakovsky2015}. However, the authors suggested that the labels provided by two human annotators did not exhibit strong overlap (one annotator's score was much lower -- around 80\%), and a significant amount of training was needed to achieve high-quality annotations.

The situation is even worse for more specialized domains, such as the diagnosis based on medical imaging, which requires years of training and experience. Moreover, due to the nature of the field, in some cases there is no clear way to classify a given observation correctly -- studies showed that medical diagnosis tests are not 100\% accurate and cannot be considered the gold standard~\cite{Joseph1995}~\cite{Bross1954}. This may be an effect of frequently occurring disagreements between medical experts interpreting test and imaging results~\cite{Bankier2010}~\cite{Mower1999}~\cite{Jarvik2009}~\cite{Carrino2009}.

Image segmentation poses an even more severe problem. Reaching an agreement whether the object of interest is present in an image is relatively easy -- what is challenging is to reach a consensus on its exact, pixel-wise location. In cases where more than one segmentation is available (which is seldom the case) there are multiple ways of handling such lack of consensus. An example of such method is the STAPLE algorithm~\cite{Warfield2004}, which automatically assigns confidence scores to each segmentation to merge multiple segmentations into one that is more accurate. 

The presence of noise in annotations may even be more pronounced in real-world datasets, which are not carefully curated and annotated. Intuition tells us that training of a deep neural network (DNN) using a dataset with non-zero annotation noise can hurt the performance of a model, since loss function calculations provide "partially incorrect" gradients, which impair the learning process. Zhu et al.~\cite{Zhu2004} investigated the effect of class label noise on the performance of a Decision Tree (DT) classifier in a classification task performed on various datasets. The study revealed that the performance of a DT classifier decays rapidly as the level of noise increases. Our recent investigation on a smaller scale (unpublished yet) revealed that classifiers based on DNNs can handle the rising amount of class label noise much better, even without applying any noise-filtering mechanisms. 

\section{Contribution}
Another very important application of computer vision, besides image classification, is image segmentation. Image segmentation is often used in medical image processing, where segmentation masks provide a visual aid for physicians. In the future, it could become the first step of automatic or semi-automatic diagnosis processes. However, we must bear in mind that the annotations provided by DL-based models are heavily dependent on the quality of the data they were trained on. Our contribution presented in this paper is three-fold: a) we show the results of our investigation of the impact of various levels of simulated noise in ground-truth segmentations on the performance of a DNN in brain tumor segmentation; b) a comparison of the DNN with a "perfect model", which learns perfectly the distribution of the simulated biases present in the data; c) the first results showing that an incorporation of bias into the loss function can partially combat a bias present in the data.

\section{Data}
\begin{figure}[ht]
\centering
\setlength{\tabcolsep}{1pt}
\begin{tabular}{lll}

a) & \multicolumn{1}{l}{b)} & \\ 
\includegraphics[width=0.23\textwidth, trim={30px 20px 30px 30px}, clip]{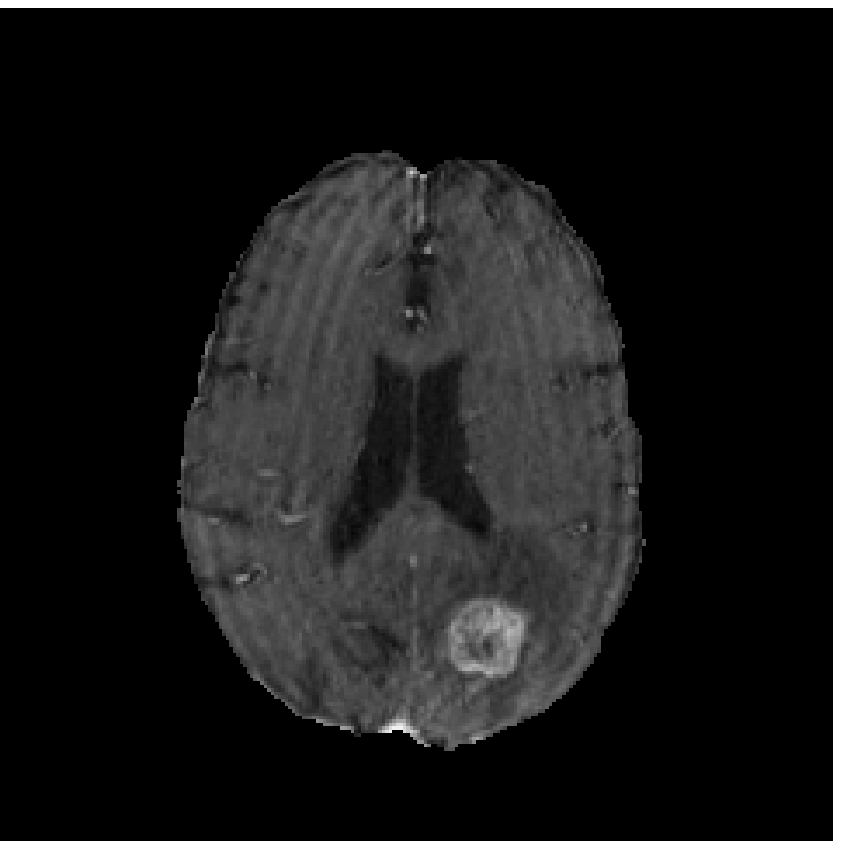} &
\includegraphics[width=0.23\textwidth, trim={30px 20px 30px 30px}, clip]{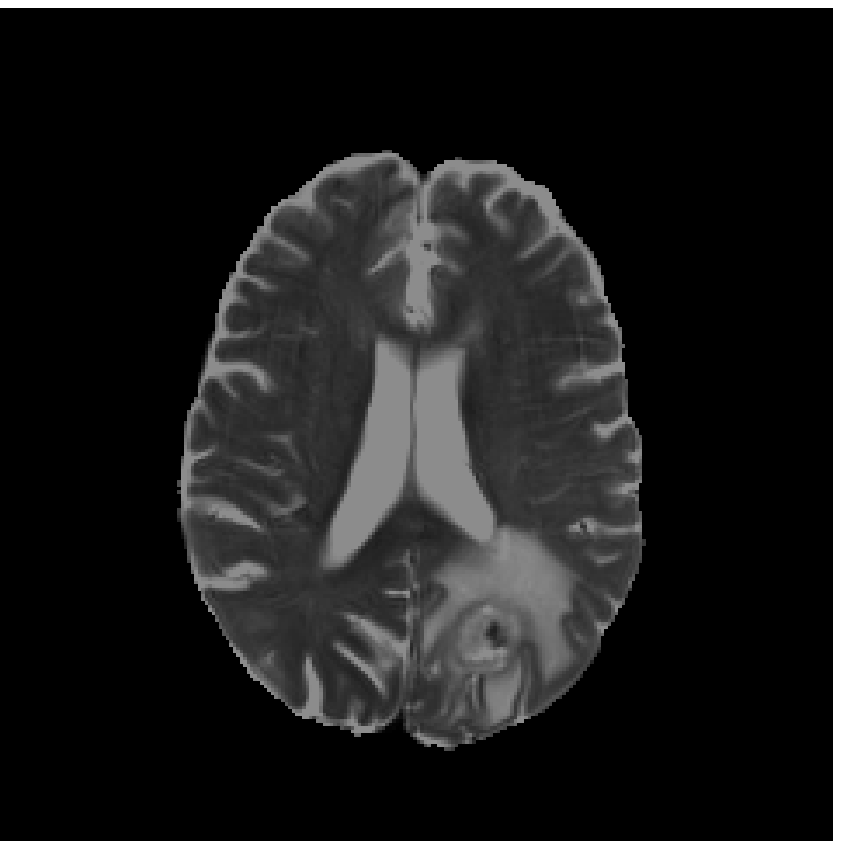}\\
c) & d) & \\ 
\includegraphics[width=0.23\textwidth, trim={30px 20px 30px 30px}, clip]{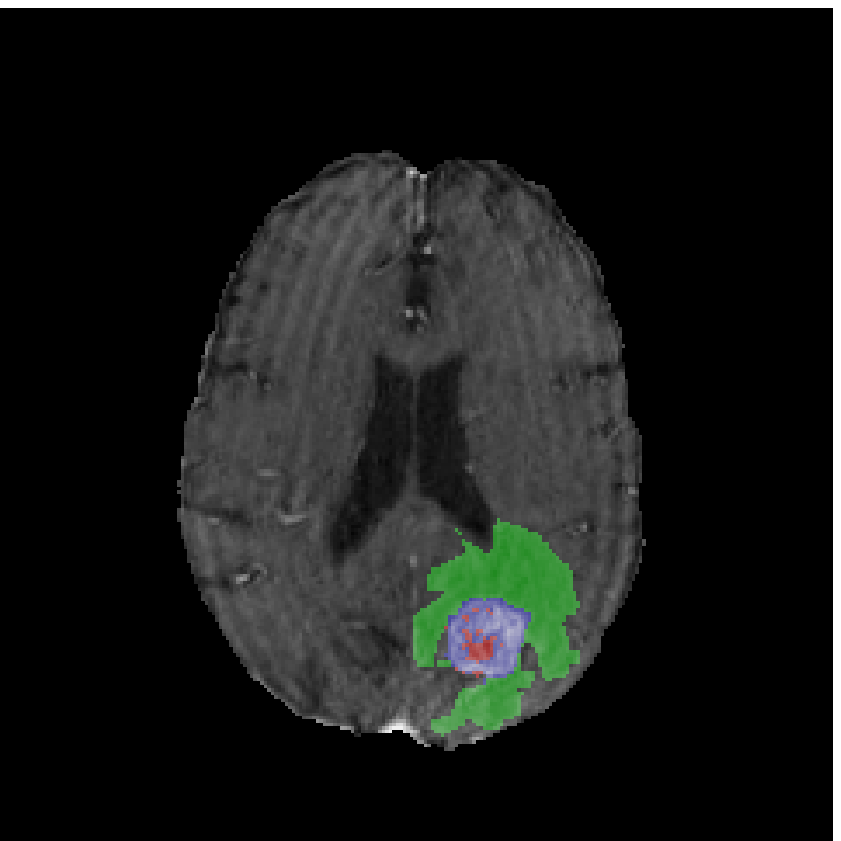} &
\includegraphics[width=0.23\textwidth, trim={30px 20px 30px 30px}, clip]{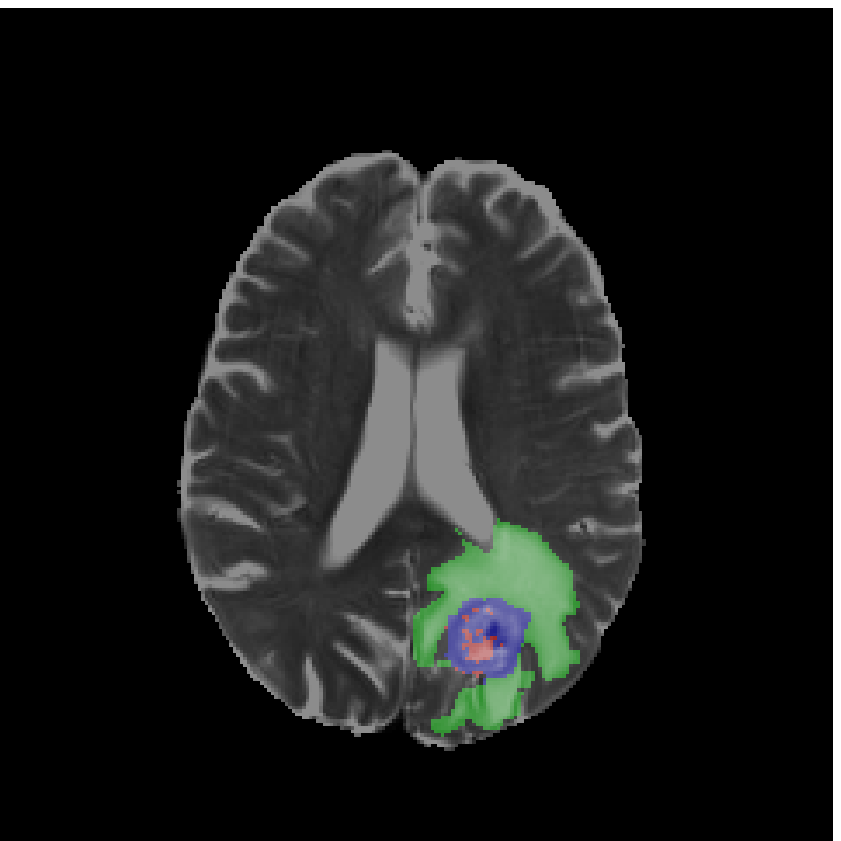}\\

\end{tabular}
\caption{Examples of images of BraTS2018 dataset in selected two modalities: a) T1c and b) T2. Their corresponding ground-truth segmentations are shown on panels c) and d), with three classes enhancing tumor (blue), peritumoral edema (green), and tumor core (red).}
\label{fig:brats2018_examples}
\end{figure}

In our study, we performed experiments on the BraTS2018 dataset~\cite{Menze2015,Bakas2017,BRATS1,BRATS2}, which consists of MRI-DCE scans of 285 patients with diagnosed gliomas: 210 patients with high-grade glioblastomas, and 75 patients with low-grade gliomas. Each study was manually labeled by one to four expert readers. The data of each patient consists of 155 frames of size 240$\times$240 px, with four co-registered modalities: native pre-contrast T1-weighted (T1), post-contrast T1-weighted (T1c), T2-weighted (T2), and Fluid Attenuated Inversion Recovery (FLAIR). The scans were skull-stripped and interpolated to the same shape (155, 240, 240) with the voxel size of 1 mm$^3$. Each pixel was assigned one of the following four labels: healthy tissue (background), Gd-enhancing tumor (ET), peritumoral edema (ED), and necrotic and non-enhancing tumor core (NCR/NET)~\cite{Bakas2017,BRATS1,BRATS2}. An example frame (T1c and T2) and the corresponding multiclass segmentation is shown in Fig~\ref{fig:brats2018_examples}. For the purpose of this work, all classes were merged into one -- whole tumor (for a binary segmentation task).

Our pre-processing followed the methodology from the BraTS2018 competition presented in~\cite{Marcinkiewicz_brats2018_proceedings2019} -- a volume-wise $z$-score normalization was applied to the brain region of each modality separately.

\section{Experiment details}
Our training was performed on a machine equipped with an Intel Core i7-7700 CPU, 64 GB RAM, and a NVIDIA GTX 1080 GPU. All experiments were performed with the PyTorch 1.0 framework in Python 3.6. In all experiments, we exploited a variation of U-net~\cite{Ronneberger2015} with residual blocks~\cite{HeResNet} consisting of just under 1M parameters. The network consisted of 3 levels with 2 residual block on contracting path (CP) and expanding path (EP), for total of 12 residual blocks. Each residual block had 3 convolutional layers with 32, 48, and 64 filters on the first, second, and third level, respectively. The data from bridge connections (used between equivalent blocks on CP and EP) was concatenated in the channel dimension with the data coming from lower level, and a single convolutional layer was used to reduce the dimensionality. Parameters of the network were optimized by a SGD optimizer with the momentum of 0.9 and initial learning rate of 0.01. The learning rate was decreased by a factor of 5 after 10 and 16 epochs. The total length of training was 20 epochs, with batch size 14 (due to memory constraints). One epoch took around 22 minutes to train. For regularization we used weight decay of $10^{-4}$.

As the main objective function, we used the dice score (\ref{eq:soft_dice}), also called the f$_1$-score, which is a harmonic mean of \textit{precision} (positive predictive value), and \textit{recall} (sensitivity). For the sake of differentiability we exploited its soft version (without thresholding). The pixel-wise dice score can be expressed as
\begin{equation}
    Dice(p, t) = \frac{2 \sum_i p_i t_i + 1.0}{\sum_i p_i + \sum_i t_i + 1.0},
\label{eq:soft_dice}
\end{equation}

where $p_i \in [0, 1]$ is the predicted value at pixel $i$, and $t_i \in \{0, 1\}$ is the target value of the same pixel, provided from the ground truth. To assure non-zero gradients and prevent division by zero, a smoothing factor of 1.0 was added to both the numerator and denominator. Since the popular DL frameworks are designed to minimize the objective function instead of maximizing it, we defined our loss function as
\begin{equation}
    \mathcal{L}(p, t) = 1.0 - Dice(p, t).
\label{eq:soft_dice_loss}
\end{equation}

The scores obtained on train and validation subsets were calculated for each frame, and then averaged; on the test subset, the scores were calculated volume-wise, which is a form of weighted-average with respect to the size of the ground-truth segmentation.

\subsection{Data split}
To validate our approach, we split the data into training, validation, and test subsets, containing 205, 40, and 40 data volumes, respectively. This allowed us to have 7 non-overlapping folds to perform cross-validation on. All the results presented are averaged over all folds.

\subsection{Simulated noise}
To imitate sub-optimal segmentations, we assumed that even expert annotators can have their own biases, and their segmentations can have a noticeable variance due to human errors. We introduced biased noise to the train and validation subsets only, since we assumed that the test subset is of sufficiently high quality to be compared against. The bias-introduction routines were based on morphological operations applied to each frame with a binary mask using a 3$\times$3 structure one or more times. The morphological operations were incorporated in three ways:

\begin{itemize}
    \item Dilate: simulates an annotator biased towards recall. The annotations produced tend to be over-segmented (the segmentations encapsulate more pixels than the true tumor), to be sure nothing important is missed. Since the tumor core is usually surrounded by the peritumoral edema, deciding exactly how far the tumor area reaches might be a non-trivial task. 
 
    \item Erode: simulates an annotator biased towards precision. The annotations produced tend to be under-segmented, ensuring that only the tumor area is included. Because of that, some parts of the tumor can be omitted. 
    
    \item Random: to simulate a random annotator or a mixture of annotators with different biases (either tending to over- or under-segment), we randomly assigned a dilation or an erosion operation for each frame in an accordingly sampled scale.
\end{itemize}

\begin{figure}[tbp]
\centering
\setlength{\tabcolsep}{1pt}
\begin{tabular}{lcccc}

a) $\Delta S = 1.00$  & 
\multicolumn{1}{l}{b) $\Delta S = 1.15$} &
\multicolumn{1}{l}{c) $\Delta S = 1.39$} & 
\multicolumn{1}{l}{d) $\Delta S = 1.61$} \\ 
\includegraphics[width=0.11\textwidth, trim={60px 0 60px 120px}, clip]{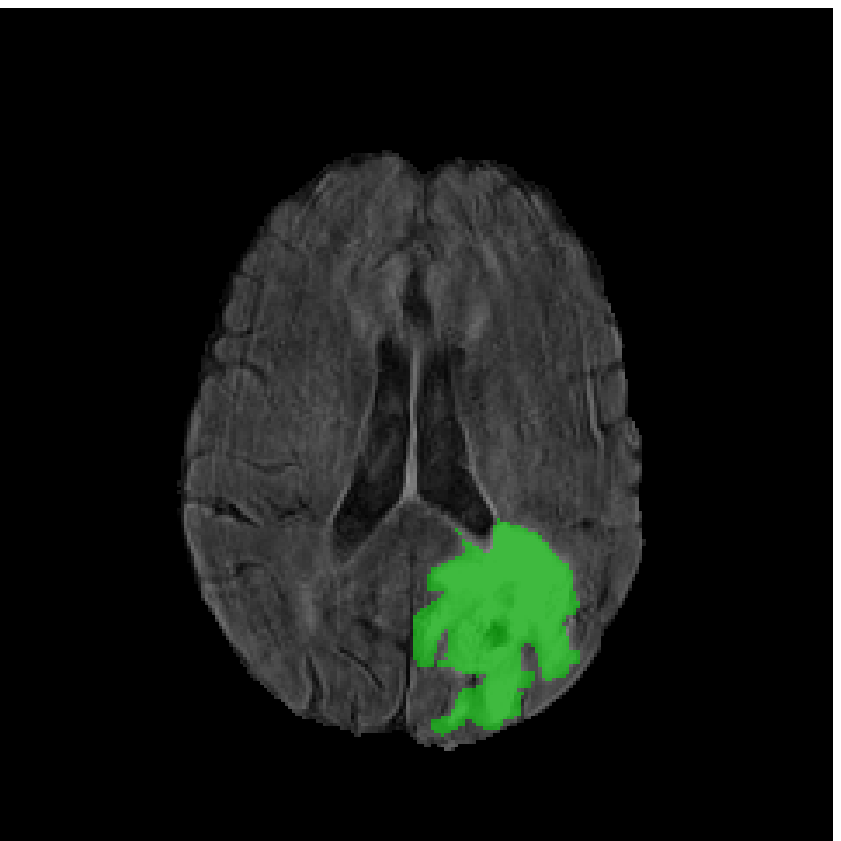} &
\includegraphics[width=0.11\textwidth, trim={60px 0 60px 120px}, clip]{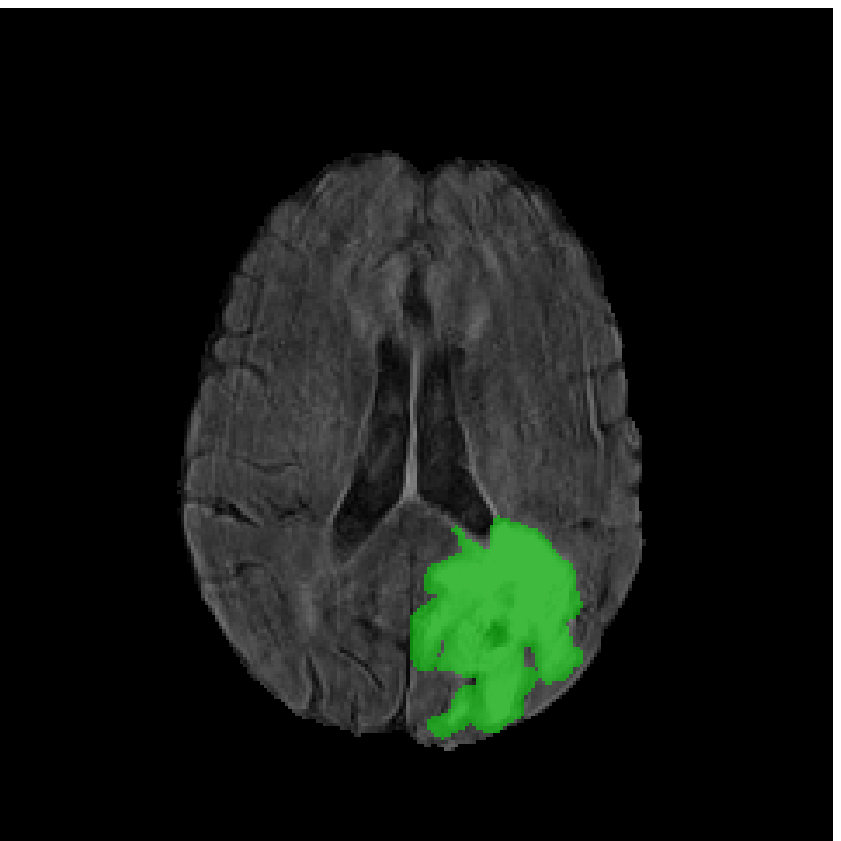} &
\includegraphics[width=0.11\textwidth, trim={60px 0 60px 120px}, clip]{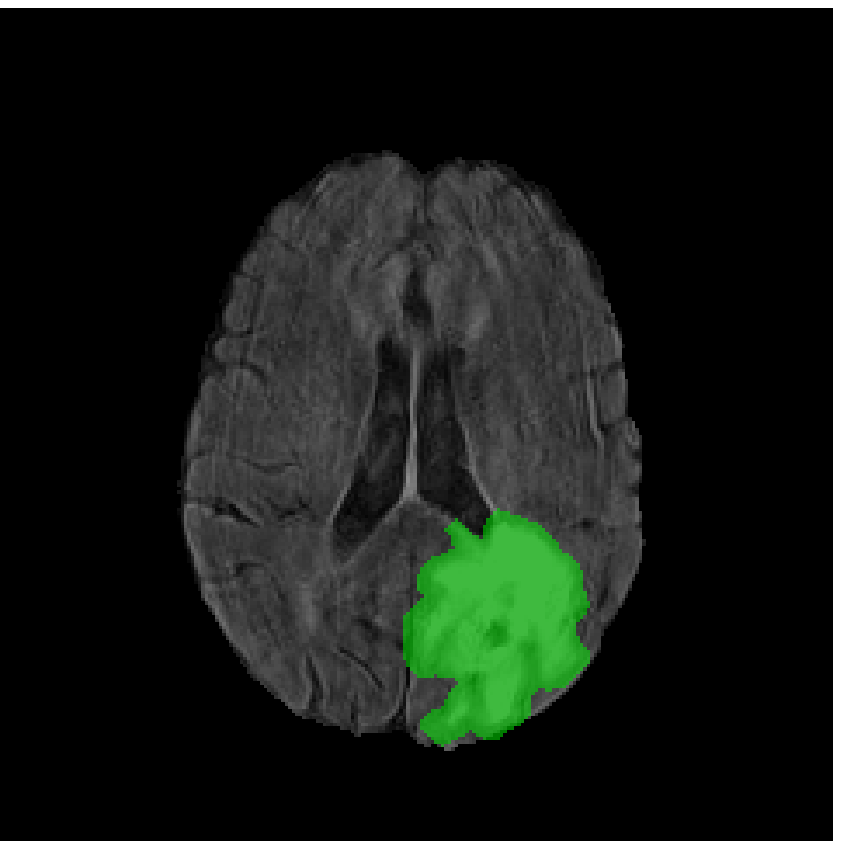} &
\includegraphics[width=0.11\textwidth, trim={60px 0 60px 120px}, clip]{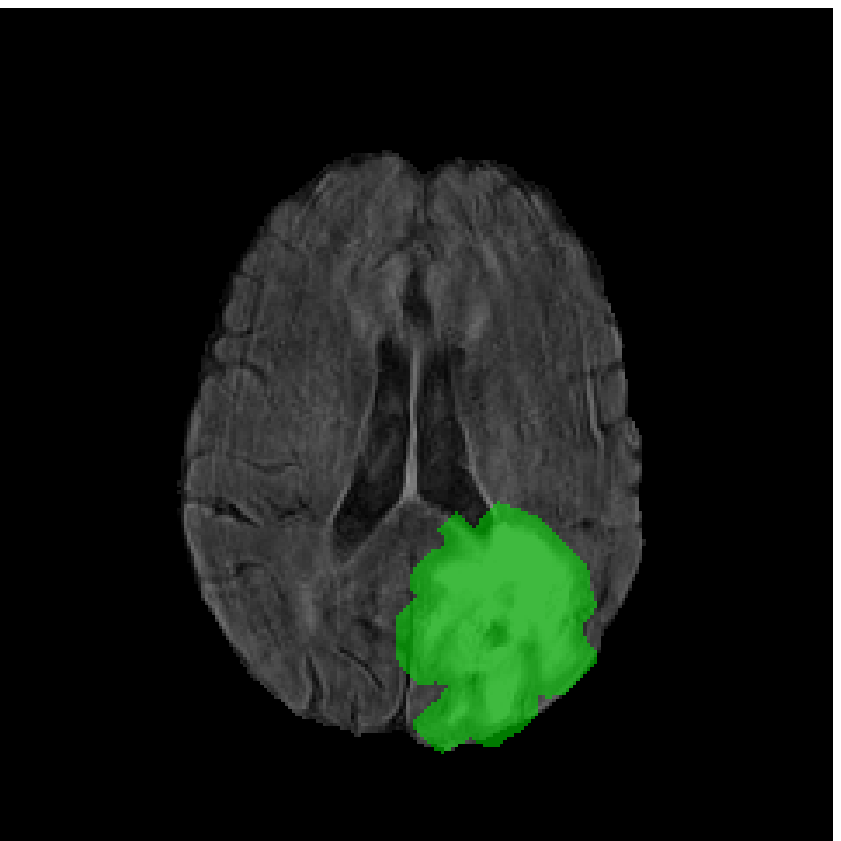}\\
e) $\Delta S = 1.00$ & 
\multicolumn{1}{l}{f) $\Delta S = 0.86$} & 
\multicolumn{1}{l}{g) $\Delta S = 0.61$} & 
\multicolumn{1}{l}{h) $\Delta S = 0.42$} \\ 
\includegraphics[width=0.11\textwidth, trim={60px 0 60px 120px}, clip]{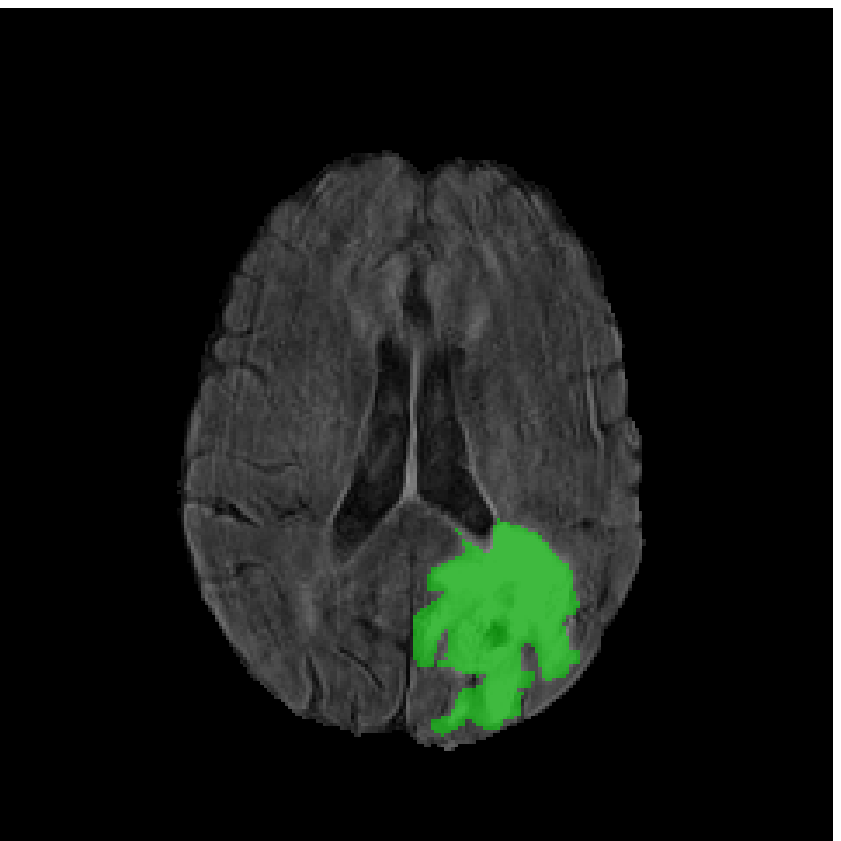} &
\includegraphics[width=0.11\textwidth, trim={60px 0 60px 120px}, clip]{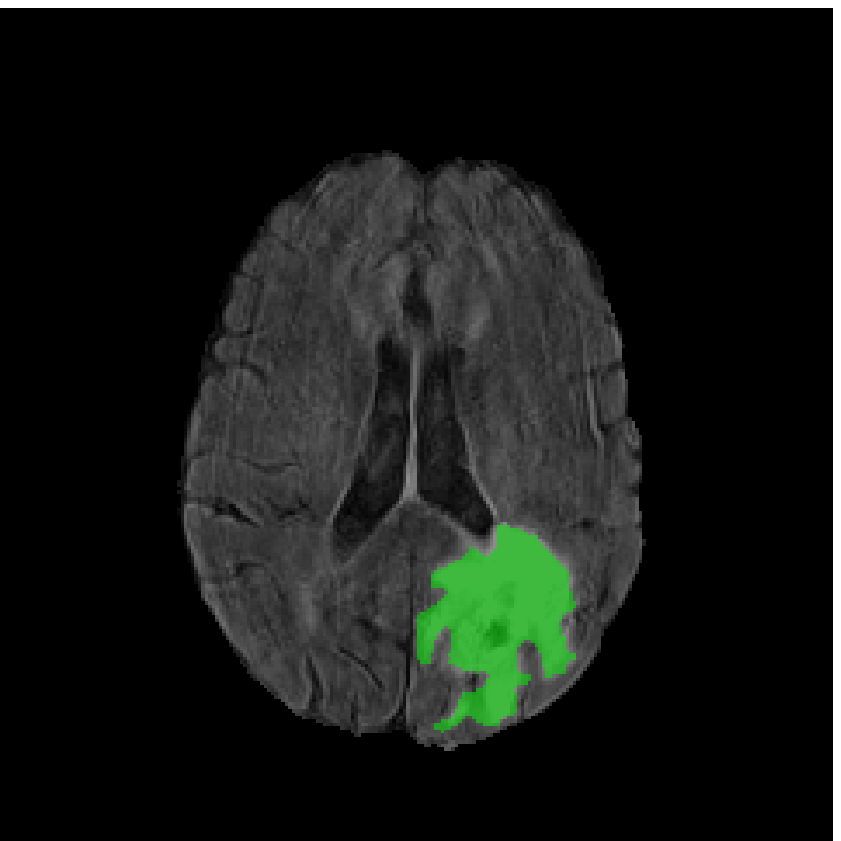} &
\includegraphics[width=0.11\textwidth, trim={60px 0 60px 120px}, clip]{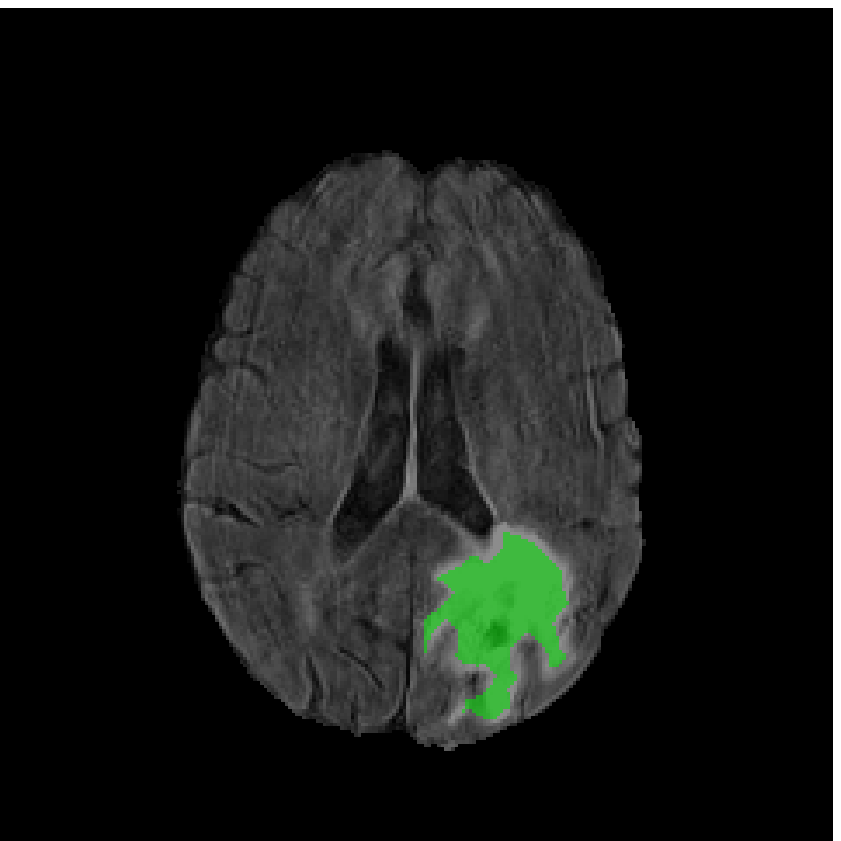} &
\includegraphics[width=0.11\textwidth, trim={60px 0 60px 120px}, clip]{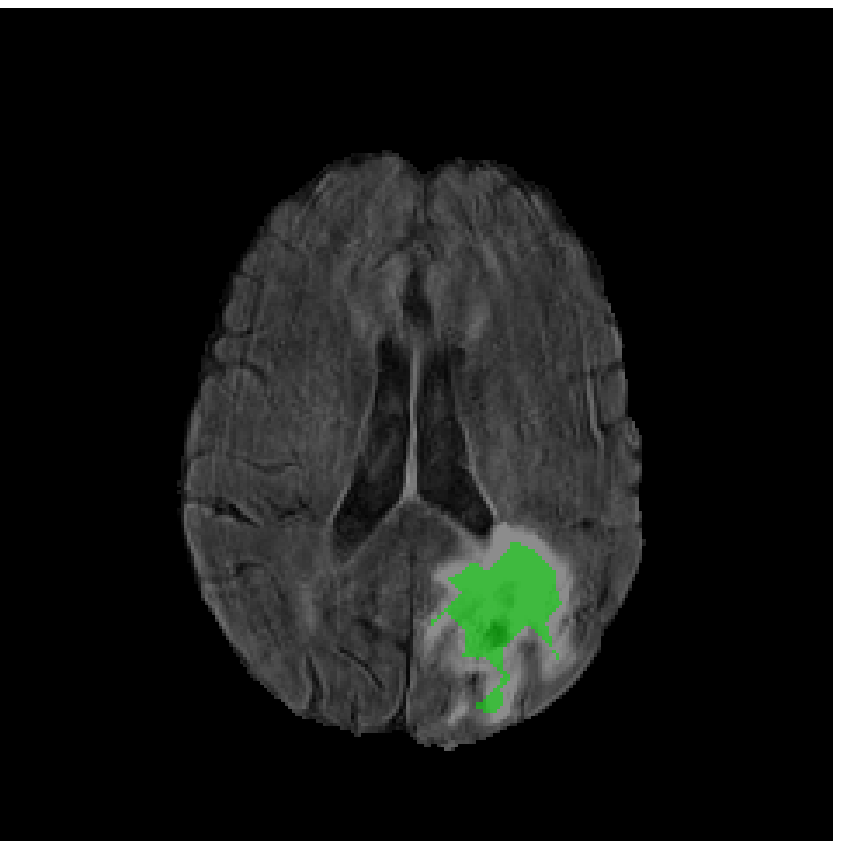}\\

Scale: 0 & 1 & 3 & 5 & \\
\hline
\end{tabular}
\caption{Examples of biased noise of a binary mask overlaid on a FLAIR image selected from BraTS2018 dataset. Original mask is presented in panels a) and e), marked by the noise scale = 0 and relative change of size $\Delta S = 1.00$. Top row (panels b, c, and d) shows examples of dilation operation with scale $\in \{1, 3, 5\}$, which translates into $\Delta S \in \{1.15, 1.39, 1.61 \}$. Bottom row (panels f, g, and h) shows examples of the erosion operation with the same scale, which translates into $\Delta S \in \{0.86, 0.61, 0.42 \}$.}
\label{fig:brats2018_morphology}
\end{figure}

The number of iterations of morphological operations, denoted here as a scale of contamination, was sampled from a normal distribution $\mathcal{N}(0, \sigma^2)$ with a few different values of variance $\sigma^2 \in \{1, 2, 3, 4, 5 \}$. Since the number of iterations had to be a positive integer number, an absolute value of the number was taken, followed by an integer casting (the \textit{floor} operation). The scale directly influenced the extent to which the original ground-truth mask was modified by a morphological operation (erosion / dilation) --- it altered the relative change of size ($\Delta S = S_{modified} / S_{original}$). If scale $=0$, the ground-truth was fed into the network unchanged, meaning that $\Delta S = 1$. Some example effects of dilation and erosion operations applied to a selected frame of FLAIR modality are presented in Fig.~\ref{fig:brats2018_morphology} for scales $\in \{0, 1, 3, 5\}$. In panels (a) and (e), the ground-truth segmentation is unchanged. The dilation operation (top panels) increased the target segmentation size by 15\%, 39\%, and 61\% for the scales of 1, 3, and 5, as shown in panels (b), (c), and (d), respectively. Erosion (bottom panels) decreased the target segmentation size by 14\%, 39\%, and 58\% for the scales of 1, 3, and 5, as shown in panels (f), (g), and (h), respectively. It is worth pointing out that the magnitude of $\Delta S$ depends strongly on the initial shape of a mask, thus morphological operations can introduce vastly different surface scaling factors.

\section{Results}
The average baseline test scores obtained by our model, without any modifications of the ground-truth segmentations, were 0.872, 0.902, and 0.863 for dice, precision, and recall, respectively. These values remained relatively consistent across all folds. The scores are comparable with some of the higher scores of the BraTS2018 challenge for the whole-tumor class on the training scoreboard. Unfortunately, since the challenge is over, we were not able to evaluate our results on the validation set or the test set, because the evaluation was carried out by the organizers. Following that, our results could not be compared with these submitted to the challenge by the participants. However, we would like to stress that multiclass segmentation (as in the BraTS2018 challenge) is a much more difficult task; the networks trained for the challenge might not have been optimized for binary segmentation, therefore there is no fair comparison between models trained for multiclass segmentation and our model. However, since our model reaches close to 0.9 of dice, precision, and recall, we are confident that it is good enough to act as a valid baseline. 

\begin{figure*}[t]
\centering
\setlength{\tabcolsep}{1pt}
\begin{tabular}{lll}

a) & b) & c)  \\ 
\includegraphics[width=0.325\textwidth]{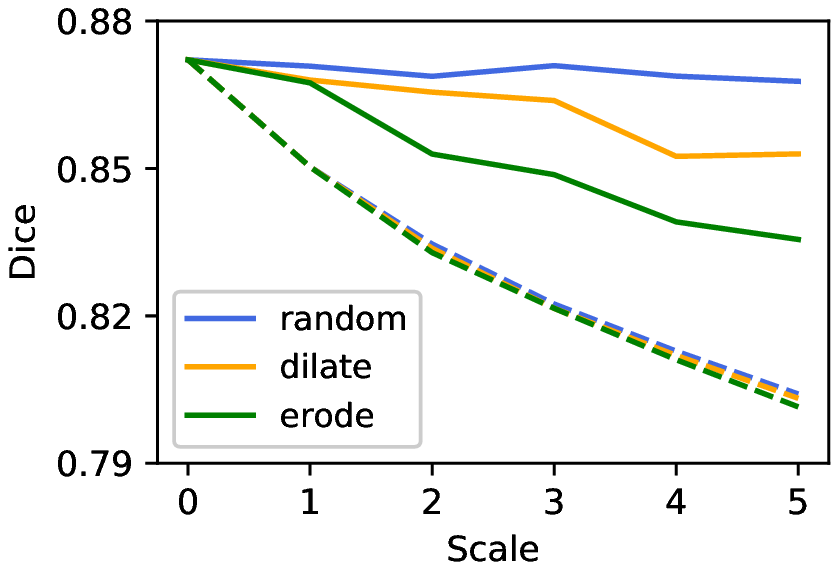} &
\includegraphics[width=0.325\textwidth]{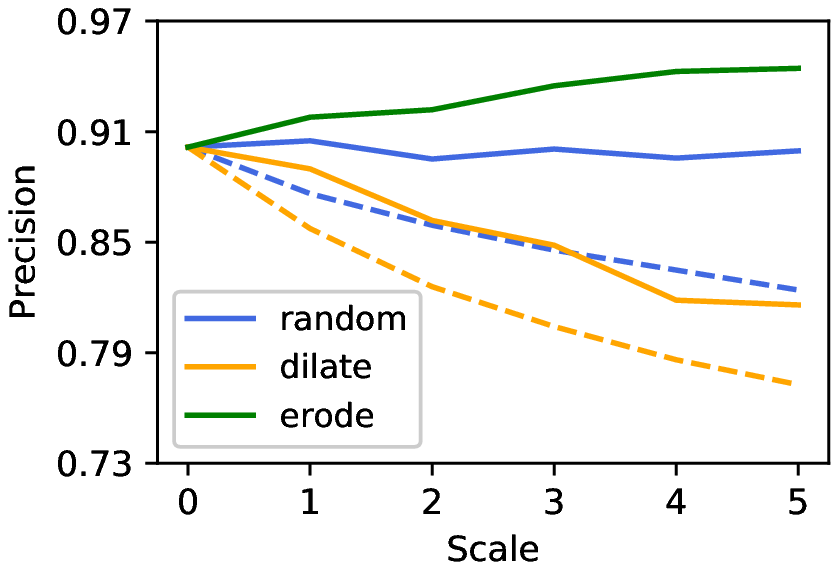} &
\includegraphics[width=0.325\textwidth]{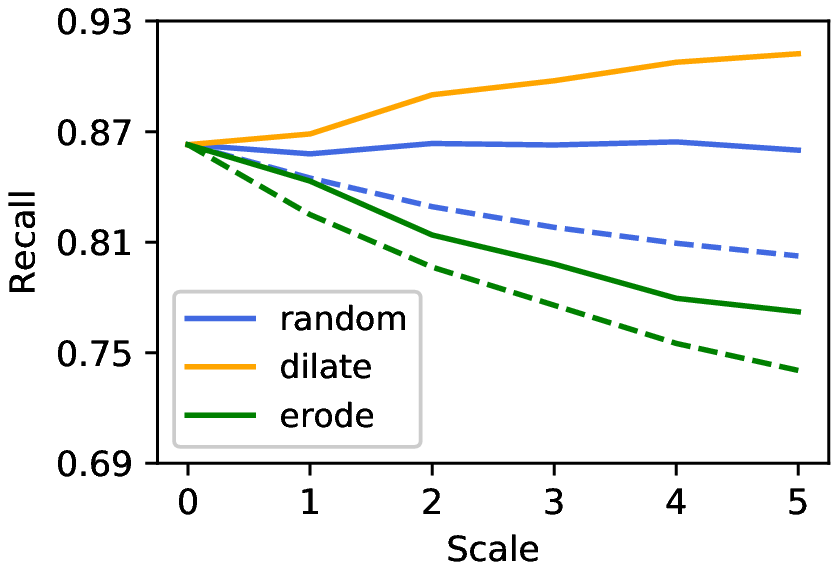}

\end{tabular}
\caption{Performance scores of a deep neural network trained using binary segmentation masks of BraTS2018 dataset with applied morphological noise simulation (erosion -- orange, dilation -- blue, or randomly chosen one -- green) as a function of the scale of the noise. Panel (a) shows the dice score, panel (b) precision, and panel (c) recall.}
\label{fig:dice_precision_recall}
\end{figure*}

\begin{table}[tbp]
\caption{Baseline dice, precision, and recall scores for our network trained and validated on BraTS2018 dataset for binary segmentation. Mean and standard deviation (std) were calculated over all folds.}
\label{tab:baseline_scores}
\centering
\begin{tabular}{|@{\vrule width0ptheight9pt\enspace}c|c|c|c|c|c|c|}\hline
\hfil & \mysplit{Val \\ Dice}  & \mysplit{Val \\ Precision} & \mysplit{Val \\ Recall} & 
\mysplit{Test \\ Dice} & \mysplit{Test \\ Precision} & \mysplit{Test \\ Recall} \\\hline

Mean & 0.896 & 0.906 & 0.880 & 0.872 & 0.902 & 0.863 \\\hline
Std & 0.013 & 0.009 & 0.021 & 0.016 & 0.020 & 0.027 \\\hline

\end{tabular}
\end{table}

The main results of this paper are presented in Fig.~\ref{fig:dice_precision_recall}. The panels (a), (b), and (c), present the dice score, precision, and recall as a function of contamination scale, respectively. Solid lines represent the results obtained by our DNN for random (blue), dilation (orange), and erosion (green) contamination modes.

We performed a simulation of a "noise-robust" model -- a model which has the same performance on noiseless data as our DNN, but also learns to mimic the noise-incorporation procedures, yielding the same performance at every scale assuming that the test set is noisy as well. Effectively, we altered the masks of each fold on each scale, and calculated all metrics against the original ground-truth segmentations. The procedure allowed us to verify how our DNN compares with the "noise-robust" model. The results for the simulated "noise-robust" model are presented with dashed lines in Fig.~\ref{fig:dice_precision_recall}. The colors match the modes of the DNN.

\subsection{Dice score} 
The dice score (Fig.~\ref{fig:dice_precision_recall}a) shows a stable behavior for random noise, degrading slightly even for higher values of scale the dice score drops only about 0.004, from 0.872 to 0.868. In the case of dilation and erosion the drop is more significant, down to 0.853 and 0.836. The results obtained by our DNN for each mode are higher than the those obtained by the simulated learner by around 8\% (random), 6\% (dilate), and 6\% (erode).

\subsection{Precision}
Random noise has a negligible effect on the precision score (Fig.~\ref{fig:dice_precision_recall}b). Erosion biases data towards precision, which is reflected in the increase of the score for that mode, from 0.902 to 0.944. Dilation has an inverse effect -- the score drops down to 0.816.  

The precision score obtained by our DNN for each mode are higher than the those obtained by the simulated learner by around 8\% (random), and 5\% (dilate). Since erosion does not misplace any pixels, the noisy mask is contained completely within the original mask -- the precision score is unaffected by such noise.

\subsection{Recall}
Random noise has similarly a negligible effect on the recall score (Fig.~\ref{fig:dice_precision_recall}c). Dilation operation favors higher recall, which is reflected by the increase of the score for that mode from 0.863 to 0.912. Contrarily, the recall score for erosion drops down to 0.772.  

The recall score obtained by our DNN for each mode are higher than the those obtained by the simulated learner by around 7\% (random), and 4\% (erode). Since dilation does not misplace any pixels, the noisy mask encapsulates completely the original mask -- the recall score is unaffected by such noise.

\subsection{Reducing bias}
We investigated whether biases present in the dataset could be proactively mitigated by altering the loss function (\ref{eq:soft_dice_loss}). We tuned the relative weight of the precision and recall (parameter $\beta$) of the dice score (\ref{eq:soft_dice}), generalizing it to the f$_{\beta}$-score, as in (\ref{eq:f_beta_score}). This operation puts more attention of the loss function towards either recall (for $\beta > 1$) or precision (for $\beta < 1$), partially countering the biases present in the data. Particularly, for $\Lim{\beta \rightarrow \infty} f_{\beta}$ yields \textit{recall}, while $\Lim{\beta \rightarrow 0} f_{\beta}$ \textit{precision}. 

\begin{equation}
    f_{\beta} = (1+\beta^2) \frac{precision \cdot recall}{\beta^2\cdot precision + recall},
\label{eq:f_beta_score}
\end{equation}

where

\begin{equation}
    precision(p, t) = \frac{\sum_i p_i t_i + 1.0}{\sum_i p_i t_i + \sum_i p_i (1-t_i) + 1.0},
\label{eq:precision_score}
\end{equation}

and

\begin{equation}
    recall(p, t) = \frac{\sum_i p_i t_i + 1.0}{\sum_i p_i t_i + \sum_i (1-p_i) t_i + 1.0}.
\label{eq:recall_score}
\end{equation}

\begin{figure}[t!]
\centering
\includegraphics[width=0.42\textwidth]{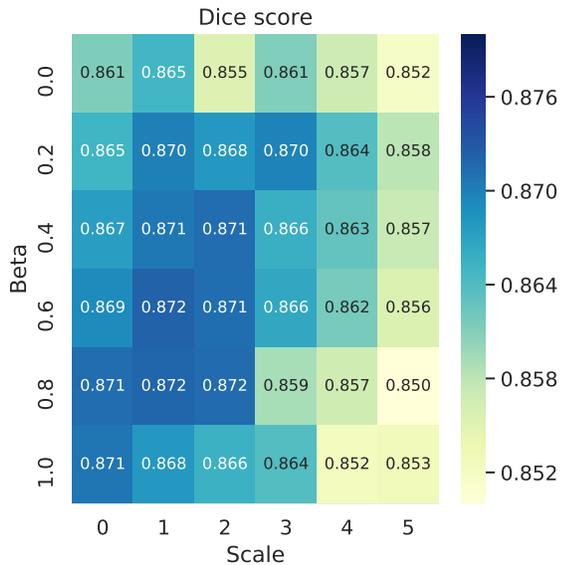}
\caption{Dice score of a deep neural network trained using binary segmentation masks of BraTS2018 dataset with applied morphological dilation as a function of the scale of the noise and a parameter \textit{beta}, representing the bias of the objective function towards precision (bias increases with decreasing \textit{beta}).}
\label{fig:beta_dice_precision_recall}
\end{figure}

To detect if the bias of the dataset could be mitigated and to what extent, we performed a gridsearch over multiple \textit{beta} values $\beta \in \{0.0, 0.2, \dots, 0.8 \}$. Lower values \textit{beta} bias the loss function towards precision, so we biased the data towards recall by using dilation. The results of the gridsearch plotted as colormaps are shown in Fig.~\ref{fig:beta_dice_precision_recall}. The values for $beta = 1.0$ were already calculated (Fig.~\ref{fig:dice_precision_recall}).

At no dilation ($scale = 0$) the dice score (Fig.~\ref{fig:beta_dice_precision_recall}a) decreases along with $beta$, which was expected as the network is no longer being trained to maximize the dice score directly. More importantly, the scores obtained for the values of $scale$ and $beta$ close to the anti-diagonal are visibly higher, especially for higher levels of noise, in comparison with the corresponding results for $\beta = 1.0$. For example, at $\beta = 0.2$, the network was able to quite consistently (for scale values $\in \{3, 4, 5\}$) score around 1.5 percent higher than for the default dice-based loss function. 

Those results confirm that indeed the effect of bias in the dataset can be offset by incorporating an opposite bias in the objective function. Most likely, the optimal performance (obtained using unbiased dataset) cannot be restored completely, but, nevertheless, the gains are non-negligible. This puts the \textit{beta} parameter as a viable hyperparameter for optimizing the performance of a deep neural network in cases where there might be a bias present in a given dataset.

\section{Conclusion}
In this paper, we investigated the impact of simulated biases and variances of annotators---reflected in the under- or over-segmentation of binary mask they annotate---on the performance of a DNN trained on such modified image-mask pairs. We employed three types of simulated modifications of original ground-truth segmentation (which we called biased noise): erosion (simulating under-segmentation bias), dilation (simulating over-segmentation bias), and random, which employed randomly either erosion or dilation. 

The results suggest that the performance of a DNN decays as the scale of contamination increases. The effect is rapid for both erosion and dilation, while it is slower (but steady) for the random contamination. This is because when training using under-segmented (eroded) segmentation masks, the DNN becomes biased towards precision, while using over-segmented (dilated) makes it biased towards recall. Both modes of contamination degrade the performance of a neural network significantly. However, for random contamination simulating a mixture of annotators with different biases, the decay of performance is less significant.

We also investigated whether the negative effect of a biased dataset on the training of a neural network could be reduced by incorporating an opposite bias in the objective function. The results confirmed that both biases partially cancel each other, thus improving the performance. We suggest that the $\beta$ parameter of the $f_{\beta}$ score be considered as an important hyperparameter to search for during the optimization. Another option worth considering is to use multiple networks trained with different values of the $\beta$ parameter in an ensemble. Such ensemble might improve the overall score via voting, just like an "ensemble" of expert annotators improve the score by improving the quality of ground-truth segmentations. 

\balance
\bibliographystyle{unsrt}
\bibliography{references}

\end{document}